\documentclass{ws-mpla}

\begin{document}

\markboth{S.I. Bastrukov, J.W. Yu, R.X. Xu and I.V. Molodtsova}
{Radiative activity of magnetic white dwarf}

%%%%%%%%%%%%%%%%%%%%% Publisher's Area please ignore %%%%%%%%%%%%%%
\catchline{}{}{}{}{}
%%%%%%%%%%%%%%%%%%%%%%%%%%%%%%%%%%%%%%%%%%%%%%%%%%%%%%%%%%%%%%%%%%%

\title{Radiative activity of magnetic white dwarf undergoing 
Lorentz-force-driven torsional vibrations}

\author{\footnotesize S.I. BASTRUKOV\footnote{Joint Institute for
 Nuclear Research,
141980 Dubna, Russia},\, J.W. YU,\, R.X. XU}

\address{State Key Laboratory of Nuclear Physics and Technology and \\
School of Physics,  Peking University, Beijing 100871, China\\
 bast@bac.pku.edu.cn}

\author{I. V. MOLODTSOVA}

\address{Joint Institute for
 Nuclear Research,
141980 Dubna, Russia
}

\maketitle

\begin{abstract}
 We study radiative activity of magnetic white dwarf undergoing torsional
 vibrations about axis of its own dipole magnetic moment under the action of Lorentz restoring force.  It is shown that pulsating white dwarf can convert
 its vibration energy into the energy of magneto-dipole emission, oscillating with the
 frequency equal to the frequency of Alfv\'en torsional vibrations, provided that
 internal magnetic field is decayed.  The most conspicuous feature of the vibration
 energy powered radiation in question is the lengthening of periods of oscillating emission; the rate of period elongation is determined by the rate magnetic field decay.
\end{abstract}


\ccode{PACS 04.40.-b.}

\section{Introduction}
 The perfect conductivity of metal-like matter of magnetic white dwarfs
 (composed of ions suspended in the Fermi-gas of relativistic electrons pervaded by
 magnetic field of a fairly high intensity, $10^5< B < 10^8$ Gauss)  suggests that
 these solid stars are capable of sustaining Alfv\'en oscillations. If this true, 
 this property should manifest itself by quasi-periodic oscillations of optical
 and X-ray emission detected form pulsating magnetic white
 dwarf stars, as was first pointed out in Ref. 1. In recent work\cite{Mol-10}, this proposition has been investigated in 
 the model of solid star with poloidal and toroidal static magnetic field with focus on
 frequency spectra of node-free regime of Alfv\'en vibrations. The main purpose of this last work was to elucidate the difference between frequency
  spectra of $a$-modes in white dwarfs having one and the same
  mass  $M$ and radius $R$, but different shapes of constant-in-time poloidal and
  toroidal magnetic fields. And it was found that each specific form of spatial
  configuration of static magnetic field is uniquely reflected in dependence of
  frequency upon overtone $\ell$ of nodeless vibrations.  Continuing this line of investigation, in this paper we explore the impact of magnetic field decay in
  pulsating magnetic white dwarf on its radiative activity.
   In section 2, a brief outline of magneto-solid-mechanical
 energy method of computing frequency of node-free Lorentz-force-driven vibrations
  of white dwarf matter in time-varying magnetic field is given. In section 3, the theory of vibration-energy
 powered magneto-dipole emission produced by magnetic white dwarf vibrating in 
 Alfv\'en mode is considered in some details.
 The obtained results are summarized in section 4.

 \section{Magnetic-field-decay effect on the energy of Alfv\'en vibrations}

 In what follows we confine our consideration to the solid-star model of magnetic (DA  
 and DB) white dwarf with poloidal (uniform and/or non-uniform internal and dipolar 
 external) magnetic field which is regarded as a slowly decreasing 
 function time. The magnetic field decay is, most likely, determined by peculiarities of 
 $\kappa$-mechanism of ionization\cite{HKT-04} (of hydrogen in DA and helium in DB stars). The time-dependent field
can be conveniently represented in the form
\begin{eqnarray}
  \label{e2.1}
 {\bf B}({\bf r},t)=B(t)\,{\bf b}({\bf r}),\quad {\bf b}({\bf r})=[b_r({\bf r})\neq 0,
  b_\theta({\bf r})\neq 0, b_\phi({\bf r})=0],
  \end{eqnarray}
 where $B(t)$ is time dependent intensity and ${\bf b}({\bf r})$ is the time-independent
 vector-function of spatial distribution of the field in the which a perfectly conducting
 matter of  volume. This leads to the following modification, compared to the case of constant in time field\cite{Mol-10,B-09a,B-09b},  of governing equations of  magneto-solid-mechanics
 \begin{eqnarray}
  \label{e2.2}
 && \rho({\bf r}){\ddot {\bf u}}({\bf r},t)=\frac{B(t)}{c}
 [\delta {\bf j}({\bf r},t)\times {\bf b}({\bf r})],\quad \nabla\cdot {\bf u}({\bf
 r},t)=0, \quad \nabla\cdot {\bf b}({\bf r})=0,\\
  \label{e2.3}
 && \delta {\bf j}({\bf r},t)=\frac{c}{4\pi}[\nabla\times \delta {\bf B}({\bf r},t)],
 \quad \delta {\bf B}({\bf r},t)=B(t)\nabla\times [{\bf u}
 ({\bf r},t)\times {\bf b}({\bf r})].
  \end{eqnarray}
 These equations describe the Lorentz-force-driven shear (non-compressional)
 differentially rotational vibrations of a perfectly conducting  solid-state plasma, regarded as non-flowing
 elastically deformable continuous medium, about axis of the time-varying magnetic
 field ${\bf B}({\bf r},t)$. In what follows we focus on the regime of node-free
  shear Alfv\'en vibrations, which has recently been extensively studied in  the context
  of a solid-star model of homogeneous density $\rho$ and
  arbitrary spatial configuration of axisymmetric internal magnetic field, ${\bf b}({\bf
  r})$. The frequency of such vibrations can be computed by using the following
  separable representation of toroidal field of differentially rotational material displacements\cite{B-07,B-08}
\begin{eqnarray}
  \label{e2.4}
 {\bf u}({\bf r},t)={\bf a}({\bf r})\,\alpha(t),\quad 
 {\bf a}({\bf r})=\nabla\times [{\bf r}\,\chi({\bf r})],\quad \nabla^2\chi({\bf r})=0.
 \end{eqnarray}
  With account of all this, the basis dynamical equation (\ref{e2.2}) can be
  represented in the form
  \begin{eqnarray}
  \label{e2.5}
 && \rho {\bf a}({\bf r}){\ddot {\alpha}}(t)=\frac{B^2(t)}{4\pi}[
 [\nabla\times[\nabla\times [{\bf a}({\bf r})\times {\bf b}({\bf r})]]]\times {\bf b}
 ({\bf r})]\,{\alpha}(t).
  \end{eqnarray}
 Scalar product of (\ref{e2.5}) with ${\bf a}({\bf r})$ and integration over the star
 leads to equation for $\alpha(t)$
 \begin{eqnarray}
  \label{e2.6}
 && {\cal M}{\ddot \alpha}(t)+{\cal K}(t)\alpha(t)=0,\quad \omega^2(t)=
 \frac{{\cal K}(t)}{\cal M},\\
   \label{e2.7}
 && {\cal M}=\rho\,m,\quad m=\int \,{\bf a}^2\,d{\cal V},\\
  \label{e2.8}
 && {\cal K}(t)=\frac{B^2(t)}{4\pi}\, k,\quad k=\int
 {\bf a}({\bf r})\cdot [{\bf b}({\bf r})\times [\nabla\times[\nabla\times [{\bf a}({\bf r})\times {\bf b}({\bf r})]]]]\,d{\cal V}.
  \end{eqnarray}
 Thus, the account of magnetic field decay leads to
 equation of non-isochronal vibrations with time-dependent spring constant ${\cal K}(t)$, contrary to harmonic in time vibrations in constant field ${\bf B}$ with constant frequency.
 It is easy to see that the magnetic field decay results in decreasing (slow-down) of
 circular frequency of vibrations
   \begin{eqnarray}
    \label{e2.9}
  && \omega(t)=\omega_A(t)\,\eta,\quad \omega_A(t)=B(t)\sqrt{\frac{R}{3M}}
  \quad\eta=\sqrt{\frac{k}{m}}\,R
  \end{eqnarray}
   and, hence, in the lengthening of the basic period of Alfv\'en vibrations
   at a rate which is determined by the rate of time-evolving suppression of magnetic field intensity $B(t)$
  \begin{eqnarray}
  \label{e2.10}
  && P_A(t)=\frac{2\pi}{B(t)}\,\sqrt{\frac{3M}{R}},\\
  \label{e2.11}
  && \frac{dP_A(t)}{dt}=-\sqrt{\frac{3M}{R}}\frac{2\pi}{B^2(t)}\frac{dB(t)}{dt}.
  \end{eqnarray}
  The most conspicuous and important feature of the vibration process under consideration
   is that magnetic field decay is accompanied by the loss of energy of Alfv\'en
   vibrations
  \begin{eqnarray}
\label{e2.12}
  && E_A(t)=\frac{{\cal M}{\dot \alpha}^2(t)}{2}+\frac{{\cal K}(t)\alpha^2(t)}{2}.
  \end{eqnarray}
  The rate of vibration energy loss
  \begin{eqnarray}
 \label{e2.13}
  &&\frac{dE_A(t)}{dt}=\frac{d{\cal K}(t)}{dt}\frac{\alpha^2(t)}{2}=
  \frac{{\cal M}\alpha^2(t)}{2}\frac{d\omega^2(t)}{dt}\\
   \label{e2.14}
  && \frac{d\omega^2(t)}{dt}=\frac{2\eta^2R^2}{3M}B(t)\frac{dB(t)}{dt}.
  \end{eqnarray}
   owes its origin to the time-evolving decay of intensity of magnetic field in the
   star.

 \section{Oscillating magneto-dipole radiation of white dwarf powered by
 energy of Alfv\'en vibrations}

We consider a case when the white dwarf converts the energy of magneto-mechanical Alfv\'en vibrations into the energy  of magneto-dipole emission. This means
that  rate of the vibration energy loss equals to the Larmor\'s  luminosity of magneto-dipole emission produced by oscillating dipole magnetic moment  $\delta \mbox{\boldmath $\mu$}(t)$ of magnetic white dwarf
 \begin{eqnarray}
 \label{e3.1}
  && \frac{dE_A(t)}{dt}=\frac{2}{3c^3}\delta {\ddot {\mbox{\boldmath
  $\mu$}}}^2(t).
  \end{eqnarray}
 The interrelation between variations in total magnetic moment $\delta \mbox{\boldmath $\mu$}(t)$ and the amplitude ${\alpha}(t)$ of 
 magneto-mechanical oscillations can be consistently interpreted provided 
 that  both above characteristics oscillate with one and the same frequency $\omega(t)$. Such a possibility is realized when
 the node-free torsional magneto-mechanical oscillations are accompanied by fluctuations of total
 magnetic moment preserving its direction: $\mbox{\boldmath $\mu$}=\mu\,{\bf
  n}={\rm constant}$. If so, $\delta {\mbox{\boldmath $\mu$}}(t)$ and
  $\alpha(t)$ must obey equations of similar form, namely
  \begin{eqnarray}
   \label{e3.2}
  && \delta {\ddot {\mbox{\boldmath $\mu$}}}(t)+\omega^2(t)
  \delta {\mbox{\boldmath $\mu$}}(t)=0,\\
   \label{e3.3}
  && {\ddot \alpha}(t)+\omega^2(t){\alpha}(t)=0,\quad \omega^2(t)=B^2(t)
  {\kappa}^2.
  \end{eqnarray}
  It follows
  \begin{eqnarray}
   \label{e3.4}
  \delta \mbox{\boldmath $\mu$}(t)=i\,\mbox{\boldmath $\mu$}\,\alpha(t),\quad i^2=-1.
  \end{eqnarray}
  On substituting  (\ref{e2.13}), (\ref{e2.14}), (\ref{e3.4}) in (\ref{e3.1}) we arrive
  at
   \begin{eqnarray}
  \label{e3.5}
  && \frac{dB(t)}{dt}=-\gamma\,B^3(t),\quad
  \gamma=\frac{2\mu^2\kappa^2}{3{\cal M}c^3}={\rm
  constant}.
  \end{eqnarray}
  As a result, we arrive at the following law of magnetic field decay 
  \begin{eqnarray}
   \label{e3.6}
  B(t)=\frac{B(0)}{\sqrt{1+t/\tau}},\quad \tau^{-1}=2\gamma B^2(0).
  \end{eqnarray}
  The lifetime of magnetic field $\tau$  is regarded as a parameter
  whose value must be established from relations
  between the period $P(t)$ and its time derivative
  ${\dot P}(t)$ which are  taken from
  observations.
  From above it follows that the time evolution of these
  latter quantities is determined by the time evolution of Alv\'en period $P_A(t)$.
  In view of this we confine our analysis to characteristic peculiarities of
  $P_A(t)$ and ${\dot P}_A$.
   
  The immediate consequence of above line of argument is the
  magnetic-field-decay induced lengthening of vibration period
 \begin{eqnarray}
  \label{e3.7}
  && P_A(t)=\frac{2\pi}{\omega_A(t)}=\frac{C_A}{B(t)},\quad
  C_A=2\pi\,\sqrt{\frac{3M}{R}}.
  \end{eqnarray}
  The rate of the Alfv\'en period elongation is given by
  \begin{eqnarray}
  \label{e3.8}
  && {\dot P}_A(t)=-\frac{C_A}{B^2(t)}\frac{dB(t)}{dt},\quad \quad \frac{dB}{dt}<0.
  \end{eqnarray}
Combining  these equations we obtain 
 \begin{eqnarray}
  \label{e3.9}
  && P_A(t)B(t)={\rm constant},\\
  \label{e3.10}
  && \frac{\dot P_A(t)}{P_A(t)}=
  -\frac{\dot B(t)}{B(t)}
  \end{eqnarray}
  These relation are 
  independent of specific form of the magnetic field decay law and, therefore,
  can be used as general argument motivating interpretation
 of quasi-periodic oscillations of optical and X-ray emission from pulsating white
 dwarf as being produced by its global Alfv\'en torsional vibrations.
  The difference between periods evaluated at successive moments of time $t_1=0$
 and $t_2=t$ is given by
 \begin{eqnarray}
 \label{e3.11}
 \Delta P_A(t)=P_A(t)-P_A(0)=-P(0)\left[1-\frac{B(0)}{B(t)}\right]>0,\quad B(t)<B(0).
 \end{eqnarray}

  \begin{figure*}
 \begin{center}
\includegraphics[width=10.cm]{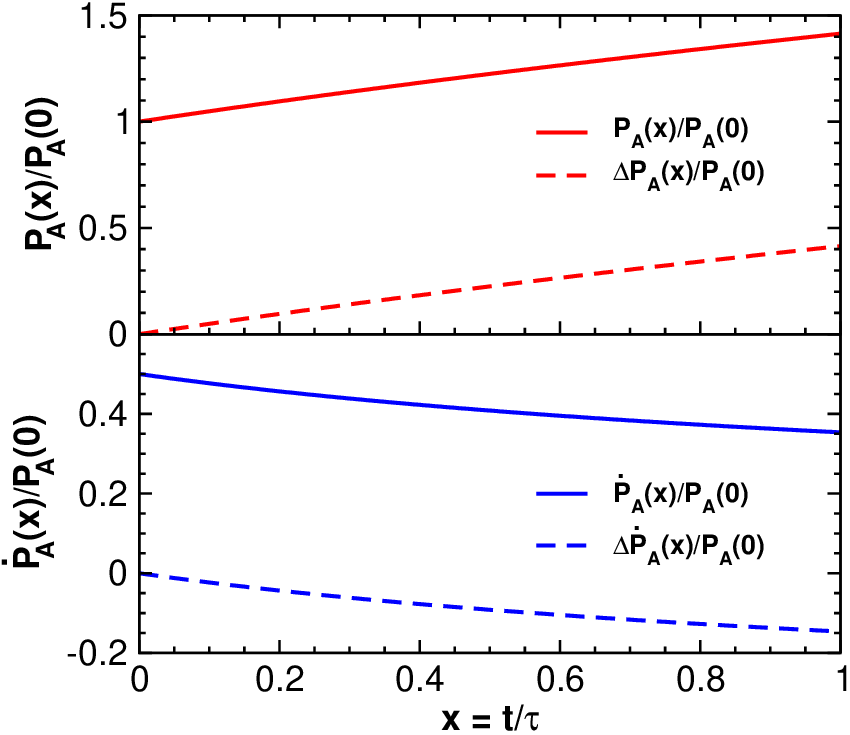}
\end{center}
\caption{
 The upper panel: the magnetic-field-decay induced elongation
 of the Alfv\'en period $P_A$ and difference between periods $\Delta P_A$ of toroidal $a$- mode, defined in the text, plotted as functions of time $t$ normalized to lifetime
 $\tau$, $x=t/\tau$. The lower panel: the time derivatives of Alfv\'en period
 ${\dot P}_A$ and difference between
 periods $\Delta {\dot P}_A$ as functions of $x$.}
 \end{figure*}

Taking into account the obtained law of magnetic field decay we obtain
    \begin{eqnarray}
  \label{e3.12}
 && P_A(t)=P_A(0)\,[1+(t/\tau)]^{1/2},\\
  \label{e3.12a}
 && {\dot P}_A(t)=\frac{1}{2\tau}\frac{P_A(0)}{[1+(t/\tau)]^{1/2}}\\
  \label{e3.14}
 && \Delta P_A(t)=P_A(0)\left[1-\sqrt{1+t/\tau}\right],\\
   \label{e3.14a}
 &&
 \Delta {\dot P}_A(t)={\dot P}_A(t)-{\dot P}_A(0)=-\frac{P_A(0)}{2\tau}\left[1-\frac{1}
 {\sqrt{1+t/\tau}}\right].
 \end{eqnarray}
 It follows that parameter of lifetime $\tau$ is given by
 \begin{eqnarray}
 \label{e3.8}
 && {\dot P}_A(t)\,P_A(t)=\frac{P^2(0)}{2\tau}={\rm constant}
 \end{eqnarray}
 and for the ratio ${\dot P}_A$ to $P_A$ we obtain
 \begin{eqnarray}
 \label{e3.9}
 &&\frac{{\dot P}_A(t)}{P_A(t)}=\frac{1}{2\tau}\,[1+(t/\tau)]^{-1}.
 \end{eqnarray}

 \begin{figure*}
 \begin{center}
 \includegraphics[width=10.cm]{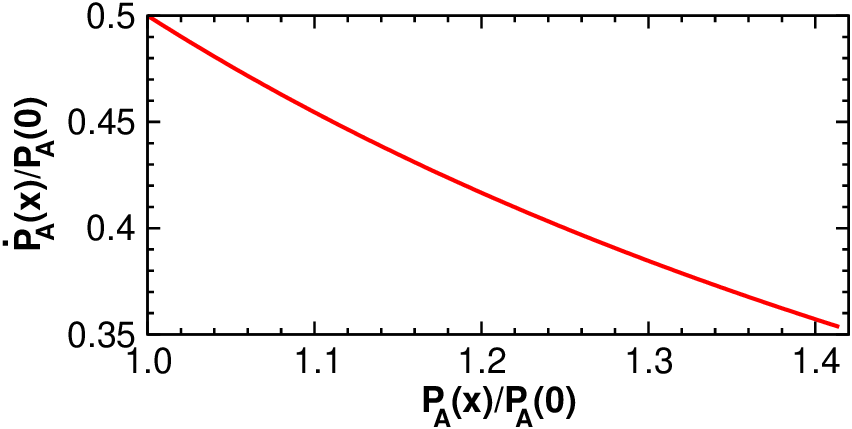}
 \end{center}
  \caption{The time derivative of period ${\dot P}_A(x)$ versus period  ${P}_A(x)$, both
 normalized to period at initial moment of time $P_A(0)$, computed as functions of
 $x=t/\tau$.}
 \end{figure*}

 In Fig.1 and Fig.2 the above quantities are plotted
 as functions of $x=t/\tau$ which is ranged in the interval $0<x<1$.

\begin{figure*}[ht]
\begin{center}
 \includegraphics[width=8.cm]{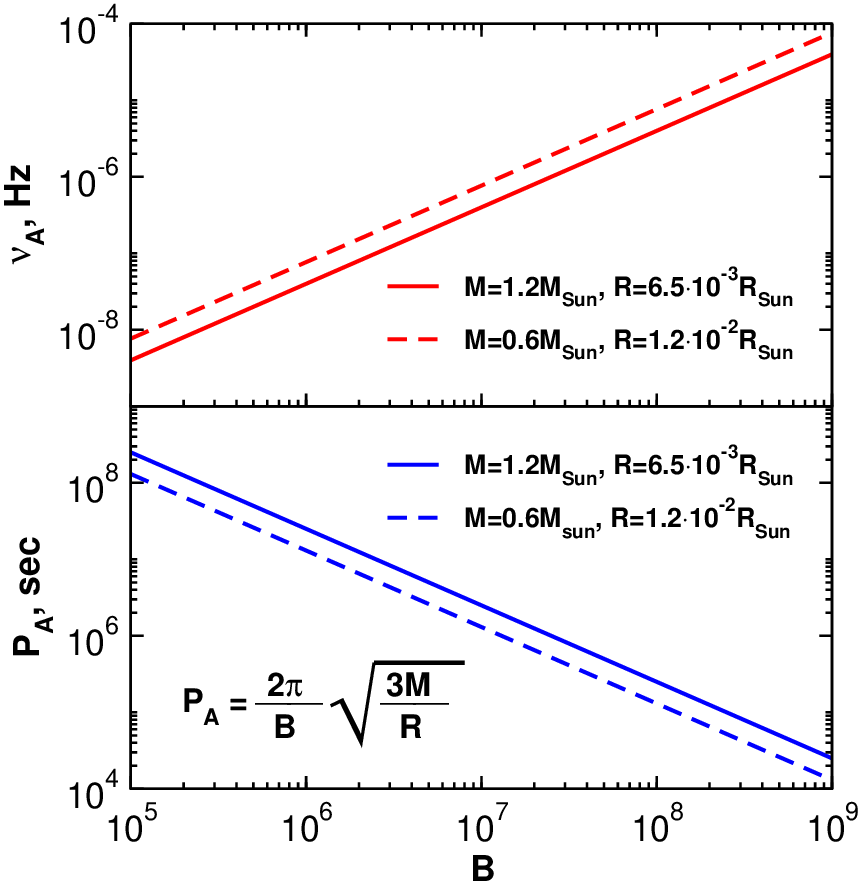}
 \end{center}
  \caption{
   Alfv\'en frequency and period of torsional magneto-mechanical 
   oscillations of white dwarfs with indicated masses and radii as functions of 
   intensity of internal magnetic field.
  }
  \end{figure*}

\section{Concluding remarks}

The most important insight to be gained from above theory
is that the magnetic field decay in magnetic white dwarf undergoing Lorentz-force-driven pulsations is inevitably accompanied by the loss of energy
of Alfv\'en oscillations of the star that causes its vibration period to lengthen at a rate proportional to the rate of magnetic field decay.  This theory rests on equations of magneto-solid mechanics lying at the base of asteroseismology of non-convective
compact objects of final stage of evolutionary track. In view of this the considered solid-star model of vibration-energy powered emission is appropriate not only for white dwarfs but also for neutron stars\cite{B-09a,B-09b}
and for strange quark stars whose material is expected to be in solid aggregate
state\cite{Xu-03}. It must be noted that the star capability
of converting the energy of Alfv\'en vibrations into energy of magneto-dipole electromagnetic emission has been considered long ago by
Hoyle, Narlikar and Wheeler\cite{HNW-64}, before the discovery of pulsars,
as one of plausible mechanisms of radiative activity of neutron stars.
The extensive consideration of such possibility  in the context of pulsating radiation from quaking magnetars is given in our recent work\cite{B-10a}. The
general conclusion of this last and the present papers is that such possibility
can be realized provided that Lorentz-force-driven Alfv\'en vibrations 
of non-convective solid star are accompanied by decay of internal magnetic field.
The characteristic frequencies and periods of such oscillations are ploted in Fig.3 for the solid-star model with typical for white dwarf masses and radii.
The predicted
elongation of vibration period should manifest itself in lengthening of period of oscillating optical and X-ray emission from white dwarf pulsating in $a$-mode 
and, hence, can be
tested by observational means.

\section*{Acknowledgment}
This work is supported by the National Natural Science Foundation of China
(Grant Nos. 10935001, 10973002), the National Basic Research Program of
China (Grant No. 2009CB824800), and the John Templeton Foundation.

\end{document}